\documentclass[conference]{IEEEtran}
\usepackage{comment}
\usepackage[final]{graphicx}
\usepackage{subfigure}
\usepackage{float}
\usepackage{amsmath}
\usepackage{cases}
\usepackage{color}
\usepackage{colortbl}
\usepackage{authblk}
\usepackage{algorithm}
\usepackage{algpseudocode}
\usepackage{amsmath}
\usepackage{amssymb}
\usepackage{booktabs}
\usepackage{setspace}
\usepackage{array}
\usepackage[a4paper, left=0.575in, right=0.575in, top=1in, bottom=1in]{geometry}

\makeatletter
\renewcommand{\maketag@@@}[1]{\hbox{\m@th\normalsize\normalfont#1}}%
\makeatother
\usepackage{stfloats}
\usepackage{cite}
\usepackage{makecell}
\usepackage{multirow}

\def\BibTeX{{\rm B\kern-.05em{\sc i\kern-.025em b}\kern-.08em
    T\kern-.1667em\lower.7ex\hbox{E}\kern-.125emX}}
    
\ifCLASSINFOpdf
\else
\fi

\usepackage{caption}
\usepackage{mathtools}

\begin{document}
\title{Delay Performance Analysis with Short Packets \\in Intelligent Machine Network}
\author[1]{Wenyan Xu}
\author[1,*]{Zhiqing Wei}
\author[2]{Zhiqun Song}
\author[1]{Yixin Zhang}
\author[1]{Haotian Liu}
\author[1]{Ying Zhou}
\author[1]{Xiaoyu Yang}
\author[2]{Yashan Pang}
\affil[1]{Beijing University of Posts and Telecommunications, Beijing 100876, China}
\affil[2]{National Key Laboratory of Advanced Communication Networks, Shijiazhuang, Hebei, P. R. China}
\affil[3]{University of Electronic Science and Technology of China, China}
\affil[1]{Email:\{wenyan.xu, weizhiqing, haotian\_liu, yixin.zhang, zhouying9705, xiaoyu.yang\} @bupt.edu.cn}
\affil[*]{Corresponding author}
\affil[2]{Email: szq\_sjz@163.com}
\affil[3]{Email: 2026201953@qq.com}

\maketitle

\begin{abstract} 
With the rapid development of delay-sensitive services happened in industrial manufacturing, Internet of Vehicles, and smart logistics, more stringent delay requirements are put forward for the intelligent machine (IM) network.
Short packet transmissions are widely adopted to reduce delay in IM networks. However, the delay performance of an IM network has not been sufficiently analyzed.
This paper applies queuing theory and stochastic geometry to construct network model and transmission model for downlink communication, respectively, proposes and derives the following three metrics, e.g., the transmission success probability (with delay as the threshold), expected delay, and delay jitter.
To accurately characterize the transmission delay with short packets, the finite blocklength capacity is used to measure the channel transmission rate.
Simulation results show that the increase of packet length and IM density significantly deteriorates the three metrics. Short packets are needed to improve the three metrics, especially in high IM density scenarios.
The outcomes of this paper provide an important theoretical basis for the optimization design and performance improvement of IM networks.

\end{abstract}

\IEEEpeerreviewmaketitle

\section{Introduction}
Intelligent machine (IM) networks are becoming increasingly important in promoting the intelligent transformation of various vertical industries, providing communication support for many delay-sensitive services\cite{9217239}\cite{feng2021joint}.
Taking industrial control services as examples, the end-to-end delay requirement for motion control is 0.5-2 ms; the delay requirement for mobile robots is 1-50 ms; and the delay requirement for remote control such as assembly robots is 4-8 ms. 
Once the delay exceeds a threshold, the control system will respond quite slowly with a decreasing control accuracy and an increasing accidental risk. To reduce latency, these applications mostly use short data packets for transmission\cite{lv2022buffer}.
In industrial control, the data packet size is usually 20-250 bits. 

Compared with long data packets, short packets occupy the channel for a shorter time, which can decrease queuing and transmission processing time, lower the probability of conflicts between multiple data packets, and thus effectively reduce the expected delay and ensure the timeliness of information~\cite{durisi2016toward}~\cite{bennis2018ultrareliable}.
On the contrary, long data packets not only results in large transmission delay, but also suffer from the unpredictable delay extension due to the network changes during the transmission process~\cite{cao2023toward}.
In multi-connection scenarios, long packets can also cause the waiting time of subsequent data packets to be extended and unstable, resulting in large delay jitter\cite{huang2024multi}. 
For example, in the data transmission of industrial automation production lines, if long packets are used to transmit machine status information, when multiple machines transmit data at the same time, queue congestion may often happen, which increases the delay jitter.
In contrast, short data packets are less affected by this kind of impact, and the delay jitter is relatively stable.
Therefore, it is important to analyze the delay performance of IM networks under finite blocklength regimes.

However, there are few studies on the theoretical derivation of IM delay performance due to many challenges.
For example, under finite blocklength conditions, the capacity formula is extremely complex owing to the factors such as channel dispersion, packet error probability, and packet length. 
In addition, due to the random spatial distribution of base stations (BSs) and IMs and the non-negligible interference between adjacent BSs, it is difficult to accurately characterize the delay performance of the IM network.

Wei \textit{et al.} in~\cite{wei2023spectrum} proposed a theoretical model for analyzing IM networks and obtained an expression of the delay performance. However, this model ignores the effect of packet length on the achievable channel rate and the obtained expression is only accurate when the packet length is infinite.
Zhang \textit{et al.} in\cite{zhang2024irs} derived a theoretical bound on the expected capacity under finite blocklength under conditions of extremely high blocking density and low transmit signal-to-noise ratio (SNR), but did not further analyze the delay performance.
Popovski \textit{et al.} in\cite{popovski2018wireless} proposed to reduce the impact of multiple BSs on delay performance through cooperation between adjacent BSs, 
but did not provide a specific theoretical analysis of multi-base station interference to explain how the BS density affects the delay performance.
Ding \textit{et al.} in~\cite{ding2024delay} analyzed the transmission delay considering both infinite and finite blocklength transmission regimes, and verified that the delay performance analysis based on the Shannon capacity is not accurate, but the paper ignored the interference in the system model and only considered the transmission delay without analyzing the end-to-end network delay performance.

To this end, we theoretically derive the expressions of three delay-related performance indicators considering finite blocklength constraints: transmission success probability, expected delay, and delay jitter, 
where the transmission success probability is defined as the probability that the transmission delay does not exceed a predefined threshold\cite{liu2024mpc}, 
the expected delay is defined as the statistically average time required for each data packet to be transmitted from the sender to the receiver, 
and delay jitter is defined as the variance of the delay\cite{gautam2021comprehensive}.
We verify the correctness of the derived expressions through simulations and study the impact of IM density and packet length on delay performance. 
Our results provide valuable insights into the design and optimization of IM systems and offer guidance for improving delay performance considering finite blocklength regimes.

The remainder of this paper is organized as follows. Section \ref{se2} presents the system model. In Section \ref{se3}, we conduct a theoretical analysis of the three delay performance metrics of IM network under short packet transmission. Section \ref{se4} details the simulation results, and Section \ref{se5} summarizes this paper.

\section{System Model}\label{se2}

\begin{figure}[!htbp]   
    \centering
    \includegraphics[width=0.48\textwidth]{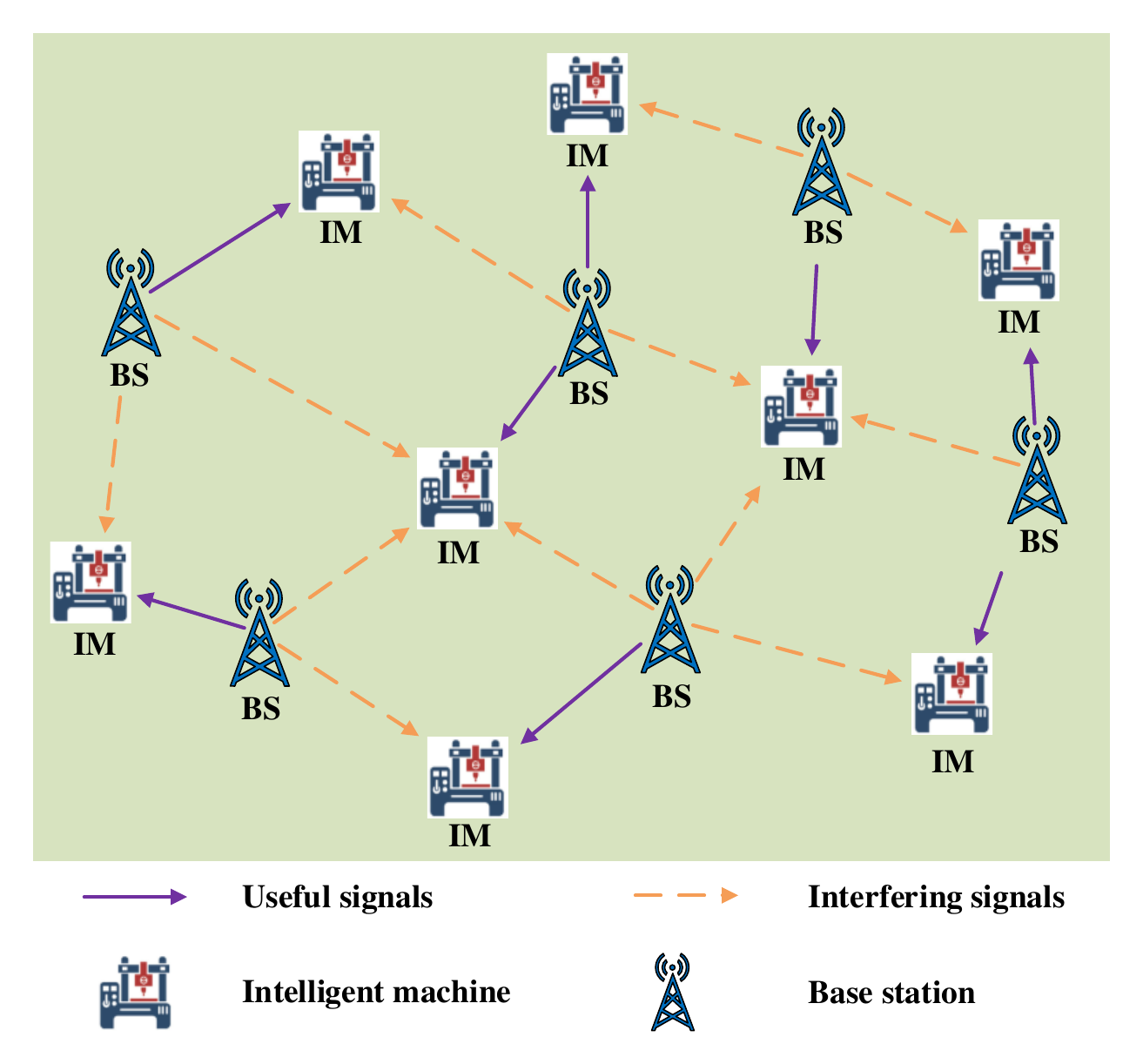}
    \caption{Downlink communication in the IM network.}
    \label{fig1}
\end{figure}

We consider downlink communication in the IM network as in Fig.~\ref{fig1}. 
Multiple BSs and $N_\text{m}$ IMs are deployed. The network is allocated a dedicated spectral bandwidth of $B_\text{m}$. 
In the context of this study, it is hypothesized
that the spectrum is evenly allocated to each IM to ensure that BSs and $N_\text{m}$ IMs can communicate simultaneously.
It is also assumed that the frequency bands used by each machine are orthogonal and there is no interference between IMs.

\vspace{-5pt}
\subsection{Network Model}
The IM network is distributed in a two-dimensional (2D) plane, and the set of BSs is denoted as ${{\phi}_\text{m}}$, whose spatial distribution follows a homogeneous poisson point process (PPP) with a density of $d_\text{m}$. The IMs are uniformly distributed within the service area of the BSs. 
According to Slyvniak's theorem~\cite{chiu2013stochastic}, the overall performance of the IM network can be evaluated by analyzing the performance of a typical IM. 

To analyze the packet processing mechanism at the BSs, an M/G/1 queuing model is employed.
For ease of analysis, we assume that packets arrive periodically and that each packet is of size $s$ bits.
The packet arrival process at the BS buffer queue is modeled as a random process that follows a Poisson distribution with an arrival rate of $\lambda$.

In the IM network, the delay experienced by each data packet consists of two components: service delay and queuing delay.
The service rate of the queue represents the average amount of information successfully transmitted per second, which is equivalent to the channel transmission rate. 
Therefore, the service time in this queuing model corresponds to the packet transmission delay $T_{\rm{t}}$, which will be further elaborated in Section \ref{se3}.

As for queuing delay, in the steady-state condition of the M/G/1 queuing model, the expected queuing delay of data packets can be expressed as\cite{ross2014introduction}
\begin{equation} \label{eq1}
    E({T_{\rm{w}}}) = \frac{{\lambda E\left( {T_{\rm{t}}^2} \right)}}{{2(1 - \lambda E\left( {T_{\rm{t}}} \right) )}},
\end{equation}
where $E\left( {T_{\rm{t}}} \right)$ represents the expected transmission delay and
$E\left( {T_{\rm{t}}^2} \right)$ represents the second-order moment of the transmission delay.

\subsection{Transmission Model} 
During packet transmission, we assume that the channel between any BS and IM experiences path loss and Rayleigh fading. All BSs transmit with the same power $P_\text{m}$. Therefore, the power of the received signal of a typical IM is ${{P_\text{m}}x_0^{ - \alpha }{h_0}}$, 
where $x_0$ represents the distance between the typical IM and the nearest BS, 
$\alpha$ represents the path loss exponent,
and $h_0$ represents the channel gain between the typical IM and the nearest BS.
While receiving the desired signal, the typical IM also receives interfering signals from other BSs. 
Therefore, the Signal to Interference plus Noise Ratio (SINR) of the typical IM received signal $\gamma _\text{m}$ is:
\begin{equation}\label{eq2}
    {\gamma _\text{m}} = \frac{{{P_\text{m}}x_0^{ - \alpha }{h_0}}}{{\sum\limits_{i \in {\phi _\text{m}}\backslash \{ 0\} } {{P_\text{m}}x_i^{ - \alpha }{h_i}} {\rm{ + }}{{N_0}}\frac{{{B_\text{m}}}}{{{N_\text{m}}}}}},
\end{equation}
where $x_i$ represents the distance between the typical IM and the $i$-th BS,  
$h_i$ represents the channel gain between the typical IM and the $i$-th BS, 
and $N_0$ is the noise power spectral density.

The formulation of (\ref{eq2}) considers the combined effects of path loss, interference, and channel noise on the SINR of a typical IM, providing a comprehensive transmission model for performance evaluation.

\section{Performance Analysis of IM Network}\label{se3}

This section focuses on theoretically analyzing three delay performance metrics under the finite blocklength regime: transmission success probability, expected delay and delay jitter.
\vspace{-10pt}
\subsection{Transmission success probability}
The transmission success probability is defined as the probability that a transmitted packet is received by the target receiver within the maximal delay limit.
Since data transmission cannot take an indefinite amount of time in practical communication scenarios, the maximum tolerable transmission delay is established within the network.
If the transmission time of a data packet exceeds the threshold, the BS is programmed to terminate the packet's transmission. 
This metric reflects the latency and reliability of IM applications, especially in delay-sensitive applications.

The packet transmission delay $T_{\rm{t}}$ in IM communication can be calculated as
\begin{equation} \label{eq3}
    {T_{\rm{t}}} = \frac{s}{R},
\end{equation}
where $R$ denotes the channel transmission rate.

The channel transmission rate in this paper cannot be simply characterized by the Shannon capacity $C$. 
Instead, it is crucial to consider the finite blocklength effect, which introduces a trade-off between reliability and achievable rate. 
For a given packet length $s$ and packet error probability $\varepsilon$, the channel transmission rate $R$ can be expressed as~\cite{5452208}
\begin{equation}  \label{eq4}
    R = C - \sqrt {\frac{V}{s}} {Q^{ - 1}}(\varepsilon),
\end{equation}
where ${Q^{ - 1}}(\varepsilon)$ is the inverse Q-function representing the reliability constraint.
$\sqrt {\frac{V}{s}} {Q^{ - 1}}(\varepsilon)$ accounts for the penalty in achievable rate due to finite blocklength effects, which becomes significant when $s$ decreases.
The channel dispersion $V$, a measure of channel variability, reflects the sensitivity of the rate to packet-length constraints.

The Shannon capacity of IM communication link ${C_\text{m}}$ is given by
\begin{equation}  \label{eq5}
    {C_\text{m}} = \frac{{{B_\text{m}}}}{{{N_\text{m}}}}{\log _2}\left( {1 + {\gamma _\text{m}}} \right),
\end{equation}
and the dispersion $V_\text{m}$ is approximated by~\cite{5452208}
\begin{equation}\label{eq6}
    {V_\text{m}} = \frac{{{\gamma _\text{m}}}}{2}\frac{{{\gamma _\text{m}} + 2}}{{{{({\gamma _\text{m}} + 1)}^2}}}{\log ^2}e,
\end{equation}
where $N_\text{m}$ denotes the number of IMs sharing the bandwidth $B_\text{m}$.

Substituting (\ref{eq5}) and (\ref{eq6}) into (\ref{eq4}), we can obtain that the channel transmission rate of the downlink communication with finite blocklength is approximated as 
\begin{equation}\label{eq7}
    {R_\text{m}} \approx \frac{{{B_\text{m}}}}{{{N_\text{m}}}}{\log _2}\left( {1 + {\gamma _\text{m}}} \right) - \sqrt {\frac{1}{{2s}}} {Q^{ - 1}}(\varepsilon )\log e.
\end{equation}

Based on (\ref{eq3}) and (\ref{eq7}), the cumulative distribution function (CDF) of transmission delay ${F_{{T_{\rm{t}}}}}\left( t \right)$ is expressed as 
\begin{equation} \label{eq8}
\begin{aligned}
    {F_{{T_{\rm{t}}}}}\left( t \right) &= P\left( {T_{\rm{t}}} < t \right)\\
    &= P\left( \gamma _\text{m} > \Theta \left( t \right)  \right)\\
    &= \exp \left( - x_0^\alpha \Theta \left( t \right) \frac{N_0 B_{\text{m}}}{N_{\text{m}} P_{\text{m}}} - d_{\text{m}}\frac{2\pi^2 x_0^2  \Theta \left( t \right) ^{\frac{2}{\alpha}}}{\alpha \sin \left( \frac{2\pi}{\alpha} \right)} \right),
\end{aligned}
\end{equation}
where 
\begin{equation} \label{eq9}
    \Theta \left( t \right) = 2^{\frac{N_{\text{m}}}{B_{\text{m}}}\frac{s}{t} + \frac{N_{\text{m}}}{B_{\text{m}}}\sqrt{\frac{1}{2s}} Q^{-1}(\varepsilon) \log e} - 1.
\end{equation}

The transmission delay threshold is set to $T_{\rm{th}}$.
Based on the definition, the transmission success probability is derived as
\begin{equation} \label{eq10}
\begin{aligned}
    {P_{\text{s}}} &= {F_{{T_{\rm{t}}}}}\left( T_{\rm{th}} \right)\\
     &= \exp \left( - x_0^\alpha  \Theta \left( T_{\rm{th}} \right) \frac{N_0 B_{\text{m}}}{N_{\text{m}} P_{\text{m}}} - d_{\text{m}}\frac{2\pi^2 x_0^2  \Theta \left( T_{\rm{th}} \right) ^{\frac{2}{\alpha}}}{\alpha \sin \left( \frac{2\pi}{\alpha} \right)} \right).
\end{aligned}    
\end{equation}
\vspace{-12pt}
\subsection{Expected delay} 
Expected delay refers to the average time it takes for each data packet to travel from the sender to the receiver, encompassing both the time required for packet transmission and the queuing delays introduced at network nodes due to congestion and scheduling constraints.

In the downlink communication link of the IM network, the expected delay $T_\text{m}$ can be calculated as
\begin{equation} \label{eq11}
    {T_{\text{m}}} = E({T_{\text{t}}}) + E({T_{\text{w}}}),
\end{equation}
where $E({T_{\text{t}}})$ can be derived as 
\begin{equation} \label{eq12}
\begin{aligned}
    E\left[ {{T_{\rm{t}}}} \right] &= E\left[ {{T_{\rm{t}}}|{T_{\rm{t}}} < {T_{{\rm{th}}}}} \right]{P_{\rm{s}}} + E\left[ {{T_{\rm{t}}}|{T_{\rm{t}}} \ge {T_{{\rm{th}}}}} \right]\left( {1 - {P_{\rm{s}}}} \right) \\
    &= {T_{{\rm{th}}}} - \int_0^{{T_{{\rm{th}}}}} {{F_{{T_{\rm{t}}}}}\left( t \right)} {\rm{d}}t,
\end{aligned}
\end{equation}

and $E({T_{\text{w}}})$ can be derived as
\begin{equation} \label{eq13}
    E\left( {{T_{\rm{w}}}} \right) = \frac{{\lambda \left( {T_{{\rm{th}}}^2 - 2\int_0^{{T_{{\rm{th}}}}} {t{F_{{T_{\rm{t}}}}}\left( t \right)} {\rm{d}}t} \right)}}{{2\left( {1 - \lambda \left( {{T_{{\rm{th}}}} - \int_0^{{T_{{\rm{th}}}}} {{F_{{T_{\rm{t}}}}}\left( t \right)} {\rm{d}}t} \right)} \right)}}.
\end{equation}

Therefore, the expected delay is expressed by
\begin{equation} \label{eq14}
\begin{aligned}
    {T_{\rm{m}}} &= {T_{{\rm{th}}}} - \int_0^{{T_{{\rm{th}}}}} {{F_{{T_t}}}\left( t \right)} {\rm{d}}t \\
    &+ \frac{{\lambda \left( {T_{{\rm{th}}}^2 - 2\int_0^{{T_{{\rm{th}}}}} {t{F_{{T_{\rm{t}}}}}\left( t \right)} {\rm{d}}t} \right)}}{{2\left( {1 - \lambda \left( {{T_{{\rm{th}}}} - \int_0^{{T_{{\rm{th}}}}} {{F_{{T_{\rm{t}}}}}\left( t \right)} {\rm{d}}t} \right)} \right)}}.
\end{aligned}   
\end{equation}

The proof is in Appendix \ref{apdA}. 

\subsection{Delay jitter}
Delay jitter measures the variability in packet delay across transmissions, which is crucial for ensuring the stability of time-sensitive applications.
High delay jitter can severely degrade the performance of IM network, particularly in applications requiring synchronized communication, such as real-time monitoring or actuation. 
In contrast, consistently low jitter is vital for maintaining predictable and stable communication.

In this paper, delay jitter is defined as the variance of the delay, which can be calculated as
\begin{equation} \label{eq15}
    {J_{\rm{m}}} = {\rm{ }}D\left( {{T_{\rm{t}}}} \right){\rm{ }} + {\rm{ }}D\left( {{T_{\rm{w}}}} \right),
\end{equation}
where $D({T_{\text{w}}})$ is the variance of queuing delay and $D({T_{\text{t}}})$ is the variance of transmission delay, which can be derived as
\begin{equation} \label{eq16}
\begin{aligned}
    D\left( {{T_{\rm{t}}}} \right) &= E(T_{\rm{t}}^2) - {\left( {E({T_{\rm{t}}})} \right)^2}\\
    &= T_{{\rm{th}}}^2 - 2\int_0^{{T_{{\rm{th}}}}} {t{F_{{T_{\rm{t}}}}}(t)} {\rm{d}}t \\
    &- {\left( {{T_{{\rm{th}}}} - \int_0^{{T_{{\rm{th}}}}} {{F_{{T_{\rm{t}}}}}(t)} {\rm{d}}t} \right)^2}.
\end{aligned}
\end{equation}

Based on (\ref{eq12}) and (\ref{eq13}), $D({T_{\text{w}}})$ can be derived as\cite{ross2014introduction}
\begin{equation} \label{eq17}
\begin{aligned}
    D\left( {{T_{\rm{w}}}} \right) &= {\left[ {E\left( {{T_{\rm{w}}}} \right)} \right]^2} + \frac{{\lambda E\left( {T_{\rm{t}}^3} \right)}}{{3\left( {1 - \lambda E\left( {{T_{\rm{t}}}} \right)} \right)}}\\
    &= {\left( {\frac{{\lambda \left( {T_{{\rm{th}}}^2 - 2\int_0^{{T_{{\rm{th}}}}} {t{F_{{T_{\rm{t}}}}}\left( t \right)} {\rm{d}}t} \right)}}{{2\left( {1 - \lambda \left( {{T_{{\rm{th}}}} - \int_0^{{T_{{\rm{th}}}}} {{F_{{T_{\rm{t}}}}}\left( t \right)} {\rm{d}}t} \right)} \right)}}} \right)^2}\\
     &+ \frac{{\lambda \left( {T_{{\rm{th}}}^3 - 3\int_0^{{T_{{\rm{th}}}}} {{t^2}{F_{{T_{\rm{t}}}}}(t)} {\rm{d}}t} \right)}}{{2\left( {1 - \lambda \left( {{T_{{\rm{th}}}} - \int_0^{{T_{{\rm{th}}}}} {{F_{{T_{\rm{t}}}}}\left( t \right)} {\rm{d}}t} \right)} \right)}},
\end{aligned} 
\end{equation}
where $E\left( {T_{\rm{t}}^3} \right)$ denotes the third-order moment of the transmission delay. 

The expression of $E\left( {T_{\rm{t}}^3} \right)$ is derived in Appendix \ref{apdA}. 

\begin{table}[!htbp]
	\centering
	\caption{Simulation Parameters
\cite{3gpp.tr.38.901.v17.0.0}\cite{3gpp.ts.22.104.2021}}
        \label{tab_simulation}
        \renewcommand{\arraystretch}{1.3} 
	\begin{tabular}{|c|c|}
		\hline
		\textbf{Parameter} & \textbf{Value} \\ \hline
		$P_{\rm{m}}$ & $24$dBm \\ \hline
		$x_0$ & $10$m \\ \hline
		$N_0$ & $10^{-10}$W/Hz  \\ \hline
		$B_{\rm{m}}$ & $100$MHz \\ \hline
		$\alpha$ & $4$ \\ \hline
		$T_{\rm{th}}$ & $0.001$s \\ \hline
	\end{tabular}
\end{table}

\section{Simulation Results}\label{se4}
We use the Monte Carlo method to verify the theoretical analysis results of Section \ref{se3}. The specific simulation architecture is as follows:
\begin{itemize}
  \item [1)] 
  Spread points. 
  According to the network model in Section \ref{se2}, BSs and IMs are regarded as random points distributed on a 2D plane. 
  The deployment of BSs and IMs is simulated by randomly scattering points in the demarcated area, where the typical IM is placed at the origin.
  Then the distances between the typical IM and BSs are calculated.  
  \item [2)]
  Generate Rayleigh fading factor. According to the channel model of Section \ref{se2}, generate random numbers that obey the Rayleigh distribution as the channel gain.
  \item [3)]
  Simulate the data packet transmission process. Generate random numbers that obey the Poisson distribution as the arrival time interval of the data packet, simulate the arrival, waiting, transmission and departure process of the data packet in the IM network, and perform statistical analysis on the delay performance of the IM network.
  \item [4)]
  Verify the correctness of the performance analysis in Section \ref{se2} by comparing with the theoretical calculation results.
\end{itemize}

The simulation parameter settings are shown in Table \ref{tab_simulation}.

\begin{figure}[htbp]
	\centering
	\subfigure[2D Perspective] {\label{fig2.a}\includegraphics[width=.45\textwidth]{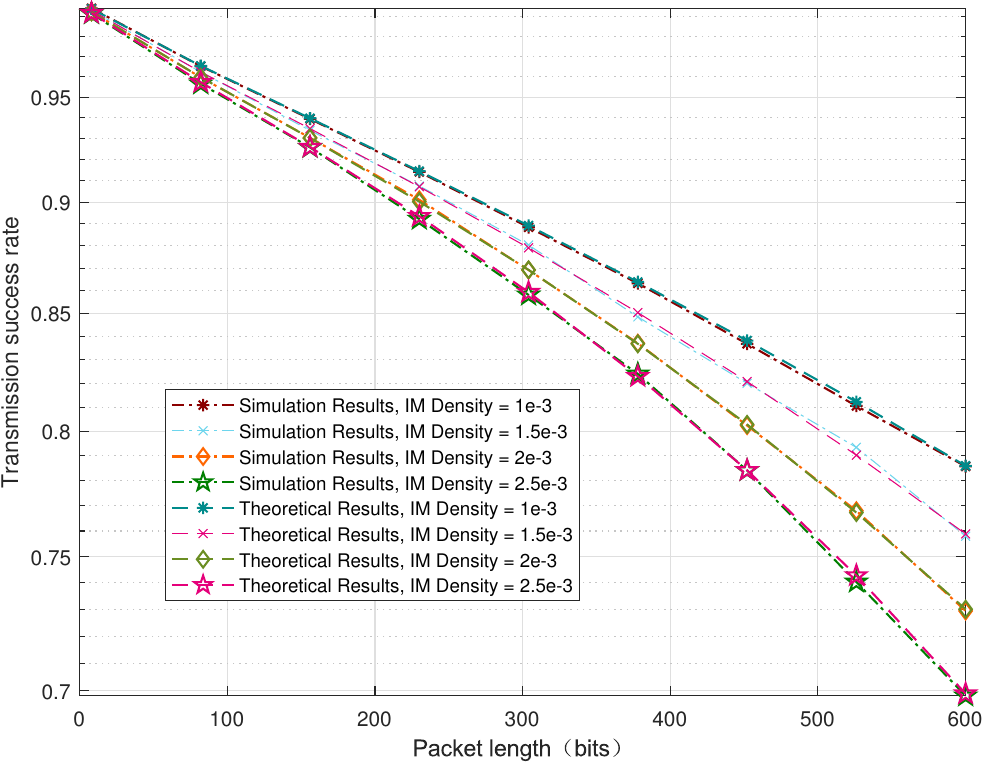}}
	\subfigure[3D Perspective] {\label{fig2.b}\includegraphics[width=.42\textwidth]{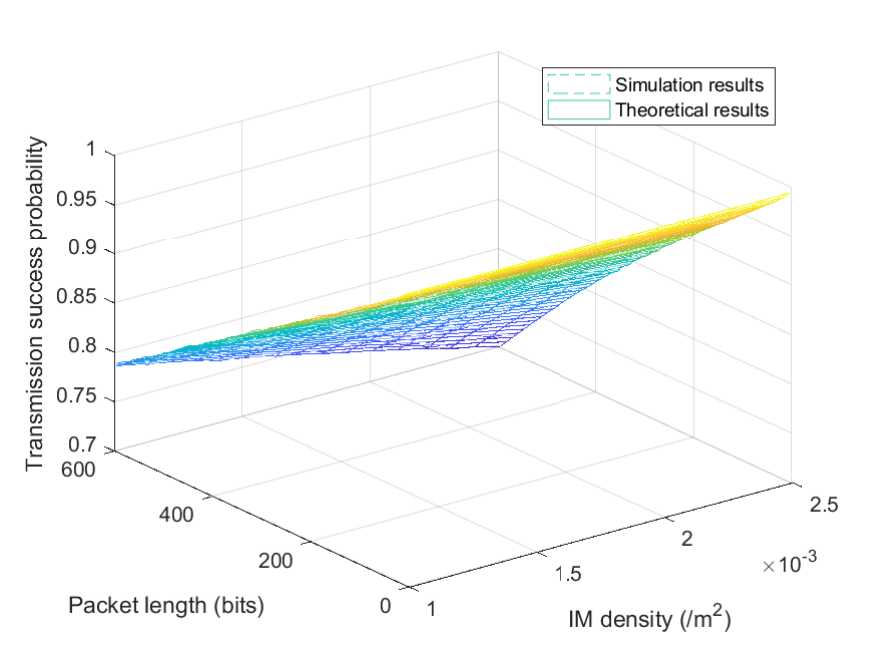}}
	\caption{Analysis of transmission success probability of IM network.}
	\label{fig.2}
\end{figure}

The impact of packet length and IM density on the transmission success probability is shown in Fig.~\ref{fig.2}. 
We give the 2D perspective (a) for a more obvious trend.
It can be seen that the theoretical results are highly consistent with the simulation results, which proves the correctness of the theoretical analysis. 
As shown in Fig.~\ref{fig.2} (b), when the packet length and IM density increase, the transmission success probability decreases significantly. When the IM density is low, the success probability is relatively high and changes more slowly with the packet length, indicating that the channel contention is low and the network performance is less affected. When the IM density is higher, the success probability decreases significantly, and is more sensitive to the change of packet length. 
This is because resource competition is fierce and the longer the data packet is, the more likely it is to fail due to delay exceeding the threshold. Therefore, in high-density scenarios, short data packets should be used preferentially to reduce delay sensitivity and improve transmission success probability.
\vspace{-7pt}
\begin{figure}[!htbp]   
    \centering
    \includegraphics[width=0.45\textwidth]{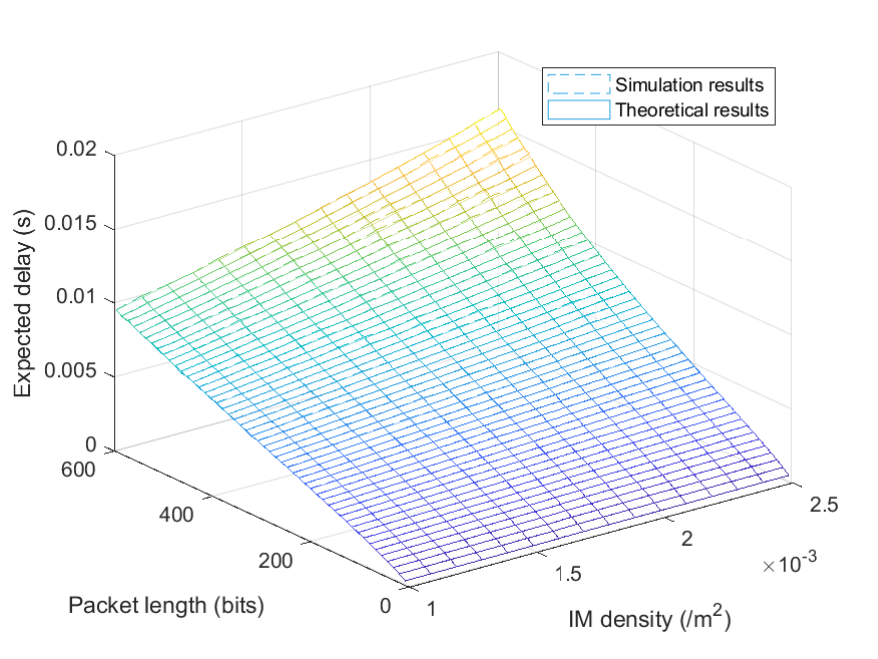}
    \caption{Analysis of delay of IM network.}
    \label{fig3}
\end{figure}
\vspace{-5pt}
Fig.~\ref{fig3} illustrates the effect of packet length and IM density on the expected delay of the IM network.
The two surfaces are consistent, which shows that the theoretical value derived from  (\ref{eq14}) for the expected delay is correct.
In addition, Fig.~\ref{fig3} shows that when the packet is small, the impact of IM density on the expected delay is relatively small. As the packet length increases, the impact gradually increases, and the inclination of the surface is significantly enhanced. 
This is because the communication rate is higher when short data packets are transmitted, so even if there is a higher density of IMs, the expected delay increase is limited.
However, long data packets need to continuously occupy more channel resources, which increases the probability of conflicts between IMs, especially in high-density networks.
At the same time, under the condition of limited blocklength, longer data packets significantly reduce the channel transmission rate according to (\ref{eq4}), thereby amplifying the impact of IM density on system delay.
In multi-BS scenario, large packets make BS scheduling and interference coordination more difficult, which further extends the expected delay.

\begin{figure}[!htbp]   
    \centering
    \includegraphics[width=0.45\textwidth]{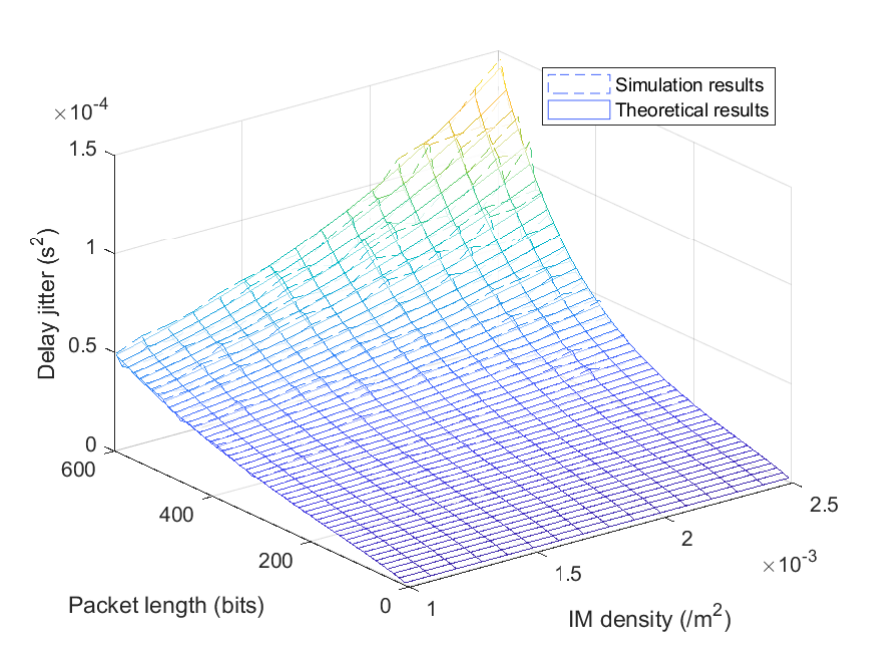}
    \caption{Analysis of delay jitter of IM network.}
    \label{fig4}
\end{figure}

Fig.~\ref{fig4} shows the impact of packet length and IM density on IM network delay jitter. 
As shown in Fig.~\ref{fig4}, the theoretical curve and the simulation curve are consistent, which shows that the theoretical solution well describes the changing trend of delay jitter. 
In addition, it can be observed that when the packet length is short, the delay jitter is small and changes smoothly. As the packet length increases, the delay jitter gradually increases, and this trend becomes more significant in high IM density scenarios. 
When the IM density is low, delay jitter remains minimal. However, as the IM density increases, delay jitter rises sharply, especially with longer packet lengths. This is because short packet transmission can significantly reduce the occupation time of channel resources, thereby mitigating delay variations caused by multi-machine resource competition.
As the IM density increases, the competition for channel resources becomes fierce, which leads to an increase in the uncertainty of delay. Especially in the case of uneven resource allocation, the delay jitter will increase significantly. Since long data packets take a long time to transmit, under high IM density, the increase in the occupancy time of a single data packet will increase network uncertainty, thereby amplifying delay jitter. 
In addition, under limited blocklength, there is a nonlinear relationship between the transmission rate and the packet error probability. Long data packets have higher requirements on latency and reliability, which further aggravates jitter.
\vspace{-6pt}
\section{Conclusion}\label{se5}
In this paper, we analyze the performance of IM networks using short packet transmission. Based on the finite packet length capacity, we propose three metrics, i.e., transmission success probability, expected delay and delay jitter, and derive their analytical expressions using stochastic geometry and queuing theory. 
The impact of IM density and packet length on the delay performance is analyzed.
Simulation results show that the obtained theoretical analysis results can accurately describe the dynamic laws of delay performance of IM network.

\begin{appendices}
\section{Derivation of transmission delay} \label{apdA}
According to the definition of $T_{{\rm{th}}}$, when ${T_{\rm{t}}} < {T_{{\rm{th}}}}$, the data packet is successfully transmitted, and the transmission delay experienced by the data packet is expressed as $T_{\rm{t}}^{{\rm{suc}}}$. 
Otherwise, the data packet fails to be transmitted, and its transmission delay is $T_{{\rm{th}}}$.
Therefore, the expected transmission delay of each packet can be calculated as
\begin{equation} \label{eq18}
\begin{aligned}
    E\left[ {{T_{\rm{t}}}} \right] &= E\left[ {{T_{\rm{t}}}|{T_{\rm{t}}} < {T_{{\rm{th}}}}} \right]{P_{\rm{s}}} + E\left[ {{T_{\rm{t}}}|{T_{\rm{t}}} \ge {T_{{\rm{th}}}}} \right]\left( {1 - {P_{\rm{s}}}} \right) \\
    &= E\left[ T_{\rm{t}}^{{\rm{suc}}} \right]{P_{\rm{s}}} + {T_{{\rm{th}}}}\left( {1 - {P_{\rm{s}}}} \right),
\end{aligned}
\end{equation}

In (\ref{eq18}), $E[T_{\rm{t}}^{suc}]$ denotes the expected value of $T_{\rm{t}}^{suc}$ for each packet, which can be derived as
\begin{equation} \label{eq19}
    E[T_{\rm{t}}^{suc}]{\rm{ = }}\int_0^{{T_{{\rm{th}}}}} {t{f_{T_{\rm{t}}^{suc}}}(t)} {\rm{d}}t,
\end{equation}
where $f_{T_{\rm{t}}^{suc}}(t)$ denotes the Probability Density Function (PDF) of $T_{\rm{t}}^{suc}$.
The Cumulative Distribution Function (CDF) of $T_{\rm{t}}^{suc}$ is defined as 
\begin{equation}\label{eq20}
    {F_{T_{\rm{t}}^{suc}}}(t) = P\left( {T_{\rm{t}}^{suc} \le t|{T_{\rm{t}}} < {T_{{\rm{th}}}}} \right).
\end{equation}

Based on the definition of conditional probability, we can deduce that
\begin{equation}\label{eq21}
\begin{aligned}
    {F_{T_{\rm{t}}^{suc}}}(t) &= \frac{{P\left( {T_{\rm{t}}^{suc} \le t,{T_{\rm{t}}} < {T_{{\rm{th}}}}} \right)}}{{P\left( {{T_{\rm{t}}} < {T_{{\rm{th}}}}} \right)}}\\
    &= \frac{{P\left( {{T_{\rm{t}}} \le t} \right)}}{{{P_{\rm{s}}}}}.
\end{aligned}
\end{equation}

Therefore, $f_{T_{\rm{t}}^{suc}}(t)$ can be expressed as
\begin{equation}\label{eq22}
\begin{aligned}
    {f_{T_{\rm{t}}^{suc}}}\left( t \right) &= \frac{{\rm{d}}}{{{\rm{d}}t}}{F_{T_{\rm{t}}^{suc}}}\left( t \right) \\
    &= \frac{{{f_{{T_{\rm{t}}}}}\left( t \right)}}{{{P_{\rm{s}}}}},
\end{aligned}
\end{equation}
where ${f_{{T_{\rm{t}}}}}\left( t \right)$ is the PDF of the packet transmission delay.

Substituting (\ref{eq22}) into (\ref{eq19}), we get the expression for $E[T_{\rm{t}}^{suc}]$ as
\begin{equation}\label{eq23}
\begin{aligned}
    E[T_{\rm{t}}^{suc}] &= \frac{1}{{{P_{\rm{s}}}}}\int_0^{{T_{{\rm{th}}}}} {t{f_{{T_{\rm{t}}}}}\left( t \right)} {\rm{d}}t\\
    &= {T_{{\rm{th}}}} - \frac{1}{{{P_{\rm{s}}}}}\int_0^{{T_{{\rm{th}}}}} {{F_{{T_{\rm{t}}}}}(t){\rm{d}}t}.
\end{aligned}
\end{equation}

Substituting (\ref{eq23}) into (\ref{eq18}), the expected transmission delay can be derived as
\begin{equation} \label{eq24}
\begin{aligned}
    E\left[ {{T_{\rm{t}}}} \right] &= E\left[ {{T_{\rm{t}}}|{T_{\rm{t}}} < {T_{{\rm{th}}}}} \right]{P_{\rm{s}}} + E\left[ {{T_{\rm{t}}}|{T_{\rm{t}}} \ge {T_{{\rm{th}}}}} \right]\left( {1 - {P_{\rm{s}}}} \right) \\
    &= {T_{{\rm{th}}}} - \int_0^{{T_{{\rm{th}}}}} {{F_{{T_{\rm{t}}}}}\left( t \right)} {\rm{d}}t.
\end{aligned}
\end{equation}

The second-order moment of the transmission delay $E\left( {T_{\rm{t}}^2} \right)$ can be calculated as
\begin{equation}\label{eq25}
    \begin{aligned}
    E\left[ {T_{\rm{t}}^2} \right] &= E\left[ {T_{\rm{t}}^2|{T_{\rm{t}}} < {T_{{\rm{th}}}}} \right]{P_{\rm{s}}} + T_{{\rm{th}}}^2\left( {1 - {P_{\rm{s}}}} \right)\\
    &= E\left[ {{{\left( {T_{\rm{t}}^{suc}} \right)}^2}} \right]{P_{\rm{s}}} + T_{{\rm{th}}}^2\left( {1 - {P_{\rm{s}}}} \right),
    \end{aligned}
\end{equation}

where $E\left[ {{{\left( {T_{\rm{t}}^{suc}} \right)}^2}} \right]$ denotes the second-order moment of $T_{\rm{t}}^{suc}$, 
which can be derived as
\begin{equation}\label{eq26}
\begin{aligned}
    E\left[ {{{\left( {T_{\rm{t}}^{suc}} \right)}^2}} \right] &= \int_0^{{T_{{\rm{th}}}}} {{t^2}{f_{T_{\rm{t}}^{suc}}}(t)} {\rm{d}}t\\
    &= T_{{\rm{th}}}^2 - \frac{1}{{{P_{\rm{s}}}}}\int_0^{{T_{{\rm{th}}}}} {2t{F_{{T_{\rm{t}}}}}(t){\rm{d}}t}. 
\end{aligned}
\end{equation}

Substituting (\ref{eq26}) into (\ref{eq25}), $E\left( {T_{\rm{t}}^2} \right)$ is derived as
\begin{equation}\label{eq27}
    E(T_{\rm{t}}^2) = T_{{\rm{th}}}^2 - 2\int_0^{{T_{{\rm{th}}}}} {t{F_{{T_{\rm{t}}}}}\left( t \right)} {\rm{d}}t.
\end{equation}

According to (\ref{eq1}), the expected queuing delay is \begin{equation} \label{eq28}
    E\left( {{T_{\rm{w}}}} \right) = \frac{{\lambda \left( {T_{{\rm{th}}}^2 - 2\int_0^{{T_{{\rm{th}}}}} {t{F_{{T_{\rm{t}}}}}\left( t \right)} {\rm{d}}t} \right)}}{{2\left( {1 - \lambda \left( {{T_{{\rm{th}}}} - \int_0^{{T_{{\rm{th}}}}} {{F_{{T_{\rm{t}}}}}\left( t \right)} {\rm{d}}t} \right)} \right)}}.
\end{equation}

Similarly, the third-order moment of the transmission delay $E\left( {T_{\rm{t}}^3} \right)$
is derived as 
\begin{equation}\label{eq29}
    E(T_{\rm{t}}^3) = T_{{\rm{th}}}^3 - 3\int_0^{{T_{{\rm{th}}}}} {t^2{F_{{T_{\rm{t}}}}}\left( t \right)} {\rm{d}}t.
\end{equation}

\end{appendices}

\vspace{+5pt}
\section*{Acknowledgment} 
This work was supported in part by the National Natural Science Foundation of China (NSFC) under Grant 92267202 and 62321001, 
in part by the Fundamental Research Funds for the Central Universities under Grant 2024ZCJH01, and in part by the National Key Laboratory of Advanced Communication Networks Foundation Project HHX24641X003.

\bibliographystyle{IEEEtran}
\bibliography{reference}

\end{document}